\documentstyle[11pt,newpasp,twoside,epsf]{article}
\def\kpc{\rm kpc}
\newcommand{\be}{\begin{equation}}
\newcommand{\ee}{\end{equation}}
\newcommand{\lab}[1]{\label{#1}}

\def\edcomment#1{\iffalse\marginpar{\raggedright\sl#1\/}\else\relax\fi}
\reversemarginpar

\begin{document}
\title{Galactic Bulge Microlensing Events with Clump Giants as Sources}
\author{P.~Popowski$^1$, C.~Alcock, R.A.~Allsman, D.R.~Alves, T.S.~Axelrod,
A.C.~Becker, D.P.~Bennett, K.H.~Cook, A.J.~Drake, K.C.~Freeman, M.~Geha,
K.~Griest, M.J.~Lehner, S.L.~Marshall, D.~Minniti, C.A.~Nelson, B.A.~Peterson,
M.R.~Pratt, P.J.~Quinn, C.W.~Stubbs, W.~Sutherland, A.B.~Tomaney, T.~Vandehei,
D.~Welch (The~MACHO~Collaboration)}
\affil{$^1$ Institute of Geophysics and Planetary Physics, Lawrence Livermore
National Laboratory; e-mail: popowski@igpp.ucllnl.org.}

\begin{abstract}
We present preliminary results of the analysis of 5 years of MACHO data
on the Galactic bulge microlensing events with clump giants as sources.
In particular, we discuss: 1) the selection of `giant' events,
2) distribution of impact parameters, 3) distribution of event durations, 
4) the concentration of long duration events
in MACHO field 104 centered on $(l,b) = (3 \fdg 1, -3 \fdg 0)$.
We report the preliminary average optical depth of 
$\tau = (2.0 \pm 0.4) \times 10^{-6} \;\;({\rm internal})$
at $(l,b) = (3\fdg 9, -3 \fdg 8)$.
We discuss future work and prospects for building a coherent
model of the Galaxy.
\end{abstract}

\section{Introduction}
The following short description of the most important observational
studies of the Galactic microlensing indicates that there is still
an urgent need for a comprehensive analysis of the microlensing
events toward the Galactic bulge.
Udalski et al.\ (1994) found 9 events in the first two year of the Optical
Gravitational Lensing Experiment (OGLE) data. They set the lower limit
on the optical depth to the Galactic bulge at $\tau = (3.3 \pm 1.2) \times
10^{-6}$. The uncertainties of this study are related to the detection
efficiency analysis as well as small number statistics.
Alcock et al. (1997) described a set of 45 events.
The potential of this sample was not fully explored due to the use of sampling
efficiencies only. The unbiased analysis was done only for 13 clump
giants, which resulted in large uncertainties of the optical depth
($\tau = 3.9^{+1.8}_{-1.2} \times 10^{-6}$).
Udalski et al.\ (2000) presented just a catalog of over 200 microlensing events
from the last 3 seasons of the OGLE-II bulge observation. Unfortunately, 
no efficiency analysis has been done for those events so the information 
that can be extracted from this sample is very limited.
Alcock et al.\ (2000a) performed Difference Image Analysis (DIA) of three
seasons of bulge data in 8 frequently sampled MACHO fields and found 99
events. They determined $\tau_{\rm bulge} = (3.2 \pm 0.5) \times 10^{-6}$.
This was a major development in bulge microlensing.
The DIA technique resulted in a substantial improvement in photometry, so
this analysis was less vulnerable to uncertainties in the parameter 
determination.
However, the results were obtained only for 8 out of 94 MACHO bulge fields.
Additionally, the detection efficiency estimate suffered from the fact
that HST luminosity function was available for only 1 field.
In summary, it is important to check the conclusions of Alcock et al.\
(2000a) with an independent set of events.

Blending is a major problem in any analysis of the 
microlensing data involving point spread function photometry.
The bulge fields are crowded, so that
the objects observed at a certain atmospheric seeing are blends of several 
stars.
At the same time, typically only one star is lensed.
In this general case, a determination of the events' parameters
and the analysis of the detection efficiency of microlensing
events is very involved and vulnerable to a number of possible systematic 
errors.
If the sources are bright one can avoid these problems. First, a 
determination of parameters of the actual microlensing events becomes
straightforward. Second, it is sufficient to estimate detection efficiency
based on the sampling of the light curve alone. This eliminates the need
of obtaining deep luminosity functions across the bulge fields. 
Red clump giants are among the brightest and most
numerous stars in the bulge.
Therefore, this analysis of the MACHO collaboration concentrates
on the events where the lensed stars are clump giants.

\section{Data}
The MACHO Project observations were performed with $1.27$-meter telescope at 
Mount Stromlo Observatory, Australia, since July 1992.
Details of the telescope system are given by Hart et al.\ (1996)
and of the camera system by Stubbs et al.\ (1993) and Marshall et al.\ (1994). 
Details of the MACHO imaging, data reduction and photometric calibration 
are described in Alcock et al.\ (1999).
In total, we collected seven season (1993-1999) of data in the 94 Galactic 
bulge fields. The bulge data that are currently available for the analysis 
consist of five seasons (1993-1997) in 77 fields (Figure 1).

\begin{figure}[h]
\epsfxsize=10cm
\centerline{\epsfbox{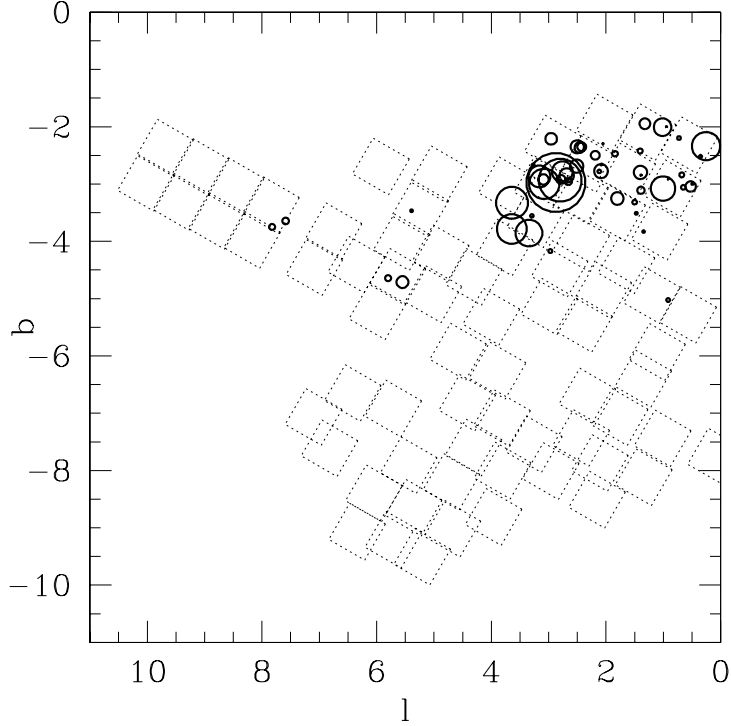}}
\caption{Location of the 77 MACHO fields [excluding 3 high-longitude fields
at $(l,b) \sim (18, -2)$] and spatial distribution of events with clump giants
as sources. The sizes of points are proportional to the Einstein ring crossing
times with the longest event \mbox{lasting $\sim 150$ days.}}
\end{figure}

\section{The Selection of Giant Events}
The events with clump giants as sources have been selected from
the sample of all events. The procedure that leads to a selection
of microlensing events of general type consists of several steps.
First, all the recognized objects in all fields are tested for any 
form of variability.
Next, a microlensing light curve is fitted to all stars showing any variation
and the objects that meet very loose selection criteria (cuts) enter the next 
phase. Here, this selection returns almost 43000 candidates.
These candidates undergo more scrutiny and are subject to more
stringent cuts, most of which test for a signal-to-noise of
the different parts of the light curve. Here, this last procedure narrows
a list of candidate events to $\sim 280$.
The question, which of those sources are clump giants, is investigated
through the analysis of the global properties of the color-magnitude
diagram (CMD) in the Galactic bulge.
The clump giant selection is based on four assumptions:\\
1) clump in the Baade's Window is representative,\\
2) the OGLE-II (e.g., Paczy\'{n}ski 1999) and MACHO photometry are
consistent,\\
3) the intrinsic colors follow the relation: $(V-R)_0 = 0.5 (V-I)_0$,\\
4) the extinction toward the bulge follows the relation: $A_V = 5.0 E(V-R)$.\\
Using the accurately measured extinction towards Baade's Window 
(Stanek 1996 with zero point corrected according to Gould et al.\ 1998 and 
Alcock et al.\ 1998) allows one to locate {\it bulge} clump giants on the 
intrinsic color -- absolute magnitude diagram. 
Such diagram can be then used to predict the positions of clump
giants on the color -- apparent magnitude diagram for fields with different
extinction.
One obtains the following average values and color ranges:
$\left< I_0^{\rm BW} \right> = 14.35$, $\left< (V-I)_0^{\rm BW} \right> = 1.1$,
$(V-I)_0^{\rm BW} \in (0.9, 1.3)$ [which corresponds to $(V-R)_0^{\rm BW} \in (0.45, 0.65)$].
Combination of average $I_0$ and $(V-I)_0$ range allows one to determine a 
central $V_0$ of the clump for a given color. For example, for $(V-I)_0=0.9$ 
one obtains $V_0=15.25$, and for $(V-I)_0$ one gets $V_0=15.65$. We 
assume that the actual clump giants scatter in $V$-mag around this 
central value, but
by not more than 0.6 mag toward both fainter and brighter $V_0$.
This defines the parallelogram-shaped box in the upper left corner
of Figure 2.

With the assumption that the clump populations in the whole bulge
have the same properties as the ones in the Baade's Window,
the parallelogram described above
can be shifted by the reddening vector
to mark the expected locations of clump giants in different fields.
The solid lines are the boundaries of the region where one could find
the clump giants of fields with different extinctions.
There are a few more $V$-mag and $(V-R)$ - color cuts that determine the
final shape of the clump region. 
The clump regions from Figure 2 contains 52 unique
clump events. There are 6 identified binary lenses among these events.

\begin{figure}
\epsfxsize=11cm
\centerline{\epsfbox{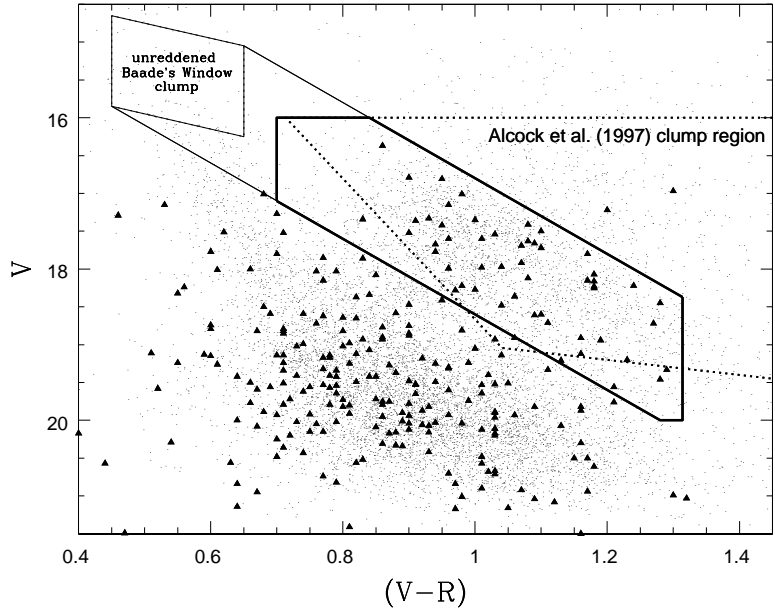}}
\caption{The region surrounded by a bold line is our clump region.
For comparison, we plot with a dotted line the clump region from Alcock et al.\ (1997). Both selections return very similar events.}
\end{figure}

Note that several assumptions that went into creating this region should be
carefully reviewed. In particular, the assumption that $(V-R)_0 = 0.5 (V-I)_0$ is only approximately true, the clump region is rather sensitive to color, 
the assumed spread in $V$ magnitudes can be either bigger or smaller
or asymmetric around the central value, clump giants in different fields may 
have different characteristics.
Taking into account that all of the above might have gone wrong, the obvious
success of the outlined procedure (see Figure 2) \mbox{is very encouraging.}

\section{Distribution of Impact Parameters}
In Figure 3, we plot the cumulative distribution of the impact parameter $u_{\rm min}$ (solid line).
The impact parameter was obtained from the maximum amplification, 
$A_{\rm max}$, according to the
formula:
\be
u_{\rm min} = \sqrt{-2 + \frac{2 \, A_{\rm max}}{\sqrt{A_{\rm max}^{2}-1}}}. 
\lab{umin}
\ee
No efficiency correction was applied.
Dashed line is the expected theoretical distribution if the minimum recorded
$A_{\rm max}$ equals to 1.5. \mbox{The agreement is beautiful.}

\begin{figure}
\epsfxsize=9cm
\centerline{\epsfbox{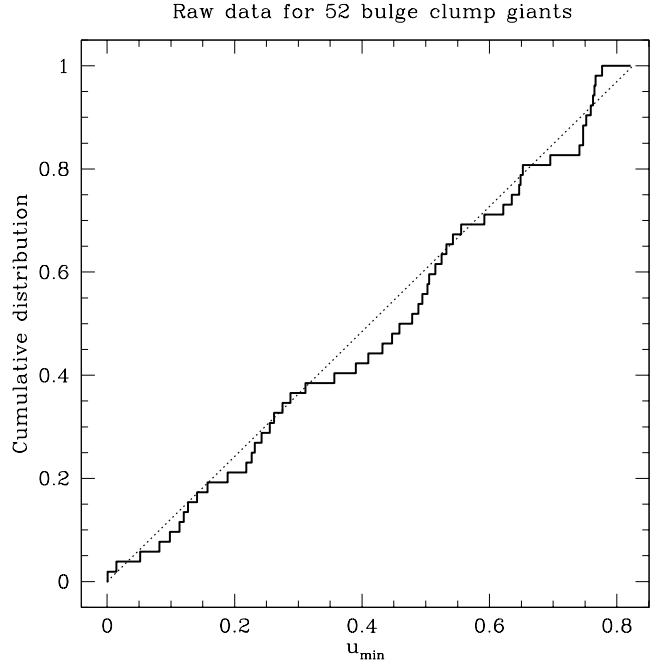}}
\caption{Cumulative distribution of impact parameters for all the candidates
including 6 binaries.}
\end{figure}

\section{Distribution of Event Durations}

We note that the first accurate description of the bulge event durations was 
given by Dr. Seuss (1960): 
``We see them come. We see them go. Some are fast. And some are slow.''
Left panel of Figure 4 shows the number of events as a function of event
duration uncorrected for detection efficiencies. The right panel presents
a contribution of events with particular duration to the optical depth.
About 40 \% of the optical depth is in the events longer than 50 days 
($t_E > 50$). This is at odds with standard models of the Galactic structure 
and kinematics.

\section{Concentration of Long Events in Field 104}

Ten clump giant events out of 52 are in the MACHO field 104.
There is a high concentration of long-duration events in this field (5 out
of 10 events longer than 50 days are in 104, including the longest 2).
We investigate how statistically significant is this concentration. 
Ideally, one would like to account for the change
in the detection of efficiency of events with different durations in
different fields.
However, the reliable efficiencies for individual fields are not available
at this point. Nevertheless, it should be possible to place a lower
limit on significance of this difference. The efficiency for detecting long
events should be similar in most fields, because this does not depend strongly 
on the sampling pattern. The detection of short events 
will be lower in a sparsely sampled fields. Therefore,
the number of short events in some of the fields used for comparison may be
relatively too small with respect to a frequently-sampled field 104, but this 
is only going to lower the significance of the $t_E$ distribution 
difference.
In conclusion, the analysis of event durations {\em uncorrected} for 
efficiencies should provide a lower limit on the difference between field 
104 and all the remaining clump giant fields.
We use the Wilcoxon's number-of-element-inversions statistic to test this 
hypothesis. 
First, we separate events into two samples: events in field 104 and all
the remaining ones.
Second, we order the event in the combined sample from the 
shortest to the longest. Then we count
how many times one would have to exchange the events from field 104 with
the others to have all the 104 field events at the beginning of the list. 
If $N_1$ and $N_2$ designate numbers
of elements in the first and second sample, respectively, then for $
N_1 \geq 4$, $N_2 
\geq 4$, and $(N_1+N_2) \geq 20$, the Wilcoxon's statistic is approximately
Gaussian distributed with an average of $N_1 N_2/2$ and a dispersion $\sigma$
of $\sqrt{N_1 \, N_2 \, (N_1+N_2+1) / 12}$.
The Wilcoxon's statistic is equal to 320,
whereas the expected number is 210 with an error of about 43. Therefore
the events in 104 differ (are longer) by $2.55 \sigma$ from the other fields.
That is, the probability that events in 104 and other fields originate from
the same parent population is of order of 0.011.

\section{The Optical Depth}

We use the following estimator of the optical depth
\be
\tau = \frac{\pi}{2NT} \sum_{\small\rm all \; events} \frac{t_E}{\epsilon(t_E)}, \lab{optdepth}
\ee
where $N$ is the number of observed stars (here about 2.1 million clump 
giants), $T$ is the total exposure (here about 2000 days) and 
$\epsilon(t_E)$ is an efficiency for detecting an event with a given $t_E$.
The sampling efficiencies were obtained with the pipeline that has been
previously applied to the LMC data (for a description see 
Alcock et al.\ 2000b). 
In brief, artificial light curves with different parameters have been
added to 1\% of all clump giants in our 77 fields and the analysis used
to select real events was applied to this set. For a given duration of the
artificial event, the efficiency was computed as a number of recovered events
divided by a number of input events.
The efficiencies used in this analysis are global efficiencies averaged over
clump giants in all 77 fields. The optical depth is reported at the
central position that is an average of positions of 1\% of the analyzed clump 
giants.\\
We obtain:
\be
\tau = (2.0 \pm 0.4) \times 10^{-6}\;\;\;\;\; {\rm at} \;\;\;\;\; (l,b) = (3 \fdg 9, -3 \fdg 8). \label{optDepth}
\ee
with the error computed according to the formula
given by Han \& Gould (1995). We caution that this result is only preliminary.
The details of the analysis as well as full discussion of the statistical 
and possible systematic errors (which may be a fair fraction of the statistical
error) will be given in Alcock et al.\ (2000c).

\begin{figure}
\epsfxsize=15cm
\centerline{\epsfbox{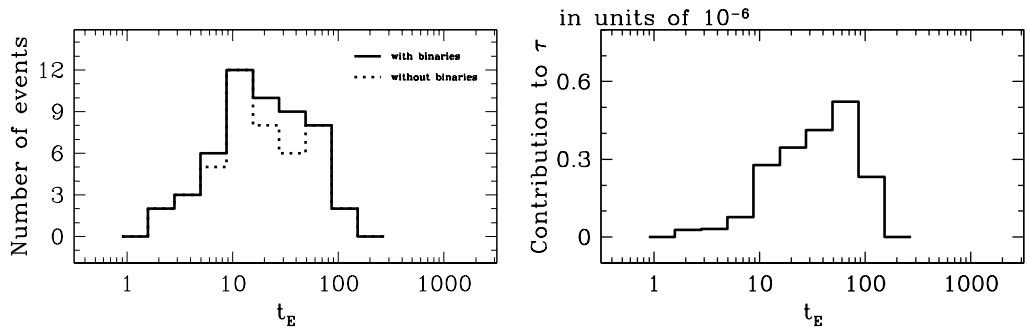}}
\caption{Left panel shows the histogram of the distribution of Einstein 
crossing times. The right panel presents the contribution of events with
different durations to the total optical depth.}
\end{figure}

\section{Future work and conclusions}
Microlensing gives two major types of constraints: the optical depth 
spatial distribution and the distribution of event durations.
The first type of information measures the mass density along different lines
of sight, whereas the second one is related to the mass function of the lenses
and the kinematics of involved populations.
These suggest two major routes of attack.
The structure of the Galaxy (e.g., disk scale length and height, bar shape
and inclination) can be constrained by analyzing the average and gradient
of the optical depth. This is most easily achieved with maximum likelihood
technique (e.g., Gyuk 1999), which allows one to extract information
from fields with and without any events.
Additionally, it is possible to constrain the location of clump sources 
comparing reddening-free indices of lensed and unlensed stars (Stanek 1995).
The difference can be up to 0.2 mag depending on the bulge/bar model, and our 
sample allows one to find the difference with the accuracy of 0.07 mag.
The microlensing data in the bulge should be combined with all the other
types of information e.g., kinematic and density distributions of RR Lyrae 
stars, rotation curve, star counts, motions of the Milky Way's 
satellites etc.

Because the distribution of event durations depends on both the mass function
of the lenses as well as kinematics of the observer, sources and deflectors,
an unambiguous solution of the entire system requires hundreds of events.
However, Han \& Gould (1996) showed that for a sample of $\sim 50$ events, 
the errors in mass function reconstruction are small if the kinematics are
assumed to be known. The kinematics are not exactly under control but it is 
possible to construct a plausible model consistent with observational 
constraints.
Determination of the mass function of the microlenses toward the Galactic bulge
would be a very exciting development. As a matter of fact, microlensing is 
currently the only method that could enable one to find a mass function of 
objects \mbox{as distant as several \kpc.}

The conclusions from the analysis described above are the following.
It is possible to select an unbiased (when $u_{\rm min}$ is concerned) 
sample of clump events in a universal way based only on $V$ and $(V-R)$ 
(or more generally, an event's position on the CMD).
Ten out of 52 clump events have durations $>50$ days, which implies that 
$\sim 40$ \% of the optical depth is in the long duration events.
This is surprising, because long events are the most likely a result
of the disk-disk lensing (Kiraga \& Paczy\'{n}ski 1994). 
However, it is widely believed that clump
giants trace the bar rather than the inner disk (Stanek et al.\ 1994).
Events in the area of field 104 centered at $(l, b) = (3 \fdg 1, -3 \fdg 0)$ 
have longer
durations than the events in other fields. The explanation can be as
exotic as a cluster of remnants along the line of sight or some conspiracy
of the bar orbits.
The optical depth averaged over the clump giants in 77 fields is
$\tau = (2.0 \pm 0.4) \times 10^{-6}$ at $(3 \fdg 9, -3 \fdg 8)$.
Allowing for the optical depth gradient of $\sim 0.5\times 10^{-6}/{\rm deg}$,
this optical depth is lower but consistent both with the 
Alcock et al.\ (1997) clump result of $3.9^{+1.8}_{-1.2}\times 10^{-6}$ at $(l,b) = 
(2 \fdg 55, -3 \fdg 64)$ as well 
as the DIA result of $(3.2 \pm 0.5) \times 10^{-6}$ at $(l,b) = (2 \fdg 68,-3 \fdg 35)$ 
(Alcock et al.\ 2000a).
The new optical depth is still rather high but in the range accessible to
\mbox{some models of the Galactic structure.}

\acknowledgments
This work was performed under the auspices of the U.S. Department of
Energy by University of California Lawrence Livermore National
Laboratory under contract No. W-7405-Eng-48.

\end{document}